\documentclass[aps,preprint,pra,epsfig,showpacs,amsmath,amsfonts,amssymb,floatfix]{revtex4-1}
\usepackage{amsmath,amsfonts,amssymb,dsfont,color,bbm,tikz,bbold}
\usepackage{graphicx}
\usepackage{dcolumn}
\usepackage{bm}
\usepackage{epstopdf}
\usepackage{epsfig}
\usepackage{environ}
\NewEnviron{myequation}{%
\begin{equation}
\scalebox{0.85}{$\BODY$}
\end{equation}
}

\begin{document}
\title{Interband coherence induced correction to Thouless pumping: Possible observation in cold-atom systems}

\author{Gudapati Naresh Raghava}
\affiliation{Department of Physics, National University of Singapore, Singapore 117546}
\author{Longwen Zhou} \email{zhoulw13@u.nus.edu}
\affiliation{Department of Physics, National University of Singapore, Singapore 117546}
\author{Jiangbin Gong} \email{phygj@nus.edu.sg}
\affiliation{Department of Physics, National University of Singapore, Singapore 117546}


\begin{abstract}
In Thouless pump, the charge transport in a one-dimensional
insulator over an adiabatic cycle is topologically quantized. For nonequilibrium
initial states, however, interband coherence will induce a previously unknown
contribution to Thouless pumping. Though not geometric in nature,
this contribution is independent of the time scale of the pumping protocol.
In this work, we perform a detailed analysis of our previous
finding [Phys. Rev. B \textbf{91}, 085420 (2015)] in an already available cold-atom setup.
We show that initial states with interband coherence can be obtained
via a quench of the system's Hamiltonian.
Adiabatic pumping in the post-quench system are then examined
both theoretically and numerically, in which the interband coherence
is shown to play an important role and can hence be observed experimentally.
By choosing adiabatic protocols with different switching-on speeds, we also show
that the contribution of interband coherence to adiabatic pumping
can be tuned. It is further proposed that the interband coherence induced correction to
Thouless pumping may be useful in capturing a topological
phase transition point. All our results have direct experimental interests.
\end{abstract}
\pacs{03.65.Vf, 05.60.Gg, 05.30.Rt, 73.20.At}


\maketitle

\section{Introduction\label{sec:Intro}}

Topological states of matter have attracted tremendous interests in
the past decades~\cite{Hasan2010,Qi2011}. A large class of
phases and phase transitions have been classified by topological
invariants~\cite{Hasan2010,Qi2011,Shen2012,Chiu2016}. One such example is the quantum
Hall effect, in which the robust quantization of Hall conductance
originates from the topological Chern number of energy bands~\cite{Thou1982}.
Other examples include topological insulators, topological superconductors,
Weyl semimetals, etc. In recent years, analogue topological states
in photonics, phononics and even acoustics have also been
realized~\cite{Lu2016,Suss2015,Yang2015}.

Most of the above mentioned topological phases are in equilibrium.
In $1983$, Thouless proposed a dynamical version of quantum Hall effect~\cite{Thou1983}.
At zero temperature, he showed that
the pumped charge through a one-dimensional insulator in an adiabatic
cycle is quantized, with the contribution from each energy band (understood as a function of the quasimomentum and time)
equal to their respective Chern numbers. At nonzero temperatures,
the pumped charge will instead be given by an integral over Berry
curvature weighted by initial populations of occupied bands. This
is an early example in which geometric and topological quantities
are introduced to describe quantum dynamics. This piece of rather fundamental physics,
now termed Thouless pump, has been demonstrated recently in two cold-atom experiments~\cite{Naka2016,Loh2016}.

In the original proposal of Thouless pump, the initial state is assumed
to be in thermal equilibrium. This is in clear contrast with recently
discovered Floquet topological phases, which are intrinsically
out-of-equilibrium~\cite{Oka2009,Inoue2010,KitagawaPRB2010,Lind2011,Jiang2011,GuPRL2011,DoraPRL2011,DerekPRL2012,KatanPRL2013,TongPRB2013,RudnerPRX2013,LongWen2014,UsajPRB2014,TorresPRL2014,DerekPRB2014,LababidiPRL2014,CarpentierPRL2015,Xiong2016,GNR2016,TitumPRX2016,MukherjeeNatCom2017}.
Moreover, there have been increasing
interests in exploring nonequilibrium response of topological states
through quantum quench~\cite{DeGottardiNJP2011,FosterPRB2013,SacramentoPRE2014,KellsPRB2014,DongNatCom2015,Aless2015,Caio2015,WangNJP2015,Caio2016,Zoller2016,Wilson2016,UnalRPA2016,Sacramento2016,HuangPRL2016,GrushinJSM2016,WangNJP2016,WangarXiv2016,BhattacharyaarXiv2016}.
In this situation,
the initial state will also be out-of-equilibrium, coherently populating
several energy bands of the post-quench Hamiltonian. These new developments
motivated us to generalize Thouless pump to situations in which nonequilibrium
initial states are prepared.

In our previous work, we developed an adiabatic perturbation theory
for periodically driven quantum systems~\cite{Hailong2015}.
Using the theory, we showed that the Thouless charge pump resulting
from a non-equilibrium initial state has two components: a weighted
integral of Floquet band Berry curvature, plus a correction due to
interband coherence in the initial state~\cite{Hailong2015}.
So long as the pumping is in the adiabatic regime, the later component is independent of the actual time scale of
pumping and can be controlled extensively via choosing different adiabatic protocols.
Later, such coherence induced correction to adiabatic pumping was found to exist in non-driven systems as well~\cite{LongWen2015},
even when the system is subject to dephasing.

In this work, taking advantage of two available
cold-atom experiments reported very recently~\cite{Naka2016,Loh2016},
we show that by suddenly quenching the cold-atom
Hamiltonian realized in the experiments, we can clearly generate the interband coherence effect in Thouless pump.
In Sec.~\ref{sec:Rice-Mele}, we introduce the already realized time-dependent
Rice-Mele model and summarize the key results of our theory. In Sec.~\ref{sec:Gaussian},
we propose to prepare initial states with interband coherence from
Gaussian wave packets through quantum quench. The resulting states
are then adiabatically pumped in parameter domains corresponding to
different topological phases.
We find that interband coherence
plays an important role in the post-quenching pumping and thus may help us to dynamically
detect and distinguish different topological phases. In Sec.~\ref{sec:Wannier},
we further test interband coherence effects by
adiabatic pumping of Wannier states after a quantum quench. In addition to leaving clear
signatures around phase transition points,
interband coherence effects are also shown to be manipulable via varying pumping protocols
and even time durations~(beyond the adiabatic regime), which suggests potential applications in quantum
control tasks. In Sec.~\ref{sec:Summary}, we summarize our results
and discuss potential future directions.

\section{Rice-Mele model and generalized Thouless pump\label{sec:Rice-Mele}}

The Rice-Mele model describes a double well lattice in one dimension~\cite{RM1982}.
Each of its unit cells has two sites
separated by a staggered onsite potential. Recently, a dynamically
modulated Rice-Mele model has been realized experimentally~\cite{Naka2016,Loh2016}.
The essential feature of this system can be described by the following
tight-binding Hamiltonian:
\begin{equation}
\hat{H}=\sum_{n=1}^{L-1}\left\{ [J+\delta(t)(-1)^{n}]c_{n}^{\dagger}c_{n+1}+{\rm h.c.}\right\} +\Delta(t)\sum_{n=1}^{L}(-1)^{n}c_{n}^{\dagger}c_{n},\label{eq:RMLattice}
\end{equation}
where $n$ is the lattice site index, and $L$ is the length of the
lattice. The distance between each pair of adjacent lattice sites
has been chosen to be $1$. Both $\delta(t)=\delta_{0}\cos(2\pi t/T)$
and $\Delta(t)=\Delta_{0}\sin(2\pi t/T)$ are periodic functions of
time $t$. Thus the tunneling amplitudes $J\pm\delta(t)$ and
energy offset $\Delta(t)$ between adjacent lattice sites are periodically
modulated in time. In momentum space, the system is described by the
Hamiltonian $\hat{H}(t)=\sum_{k}|k\rangle\langle k|h(k,t)$, with
\begin{equation}
H(k,t)=\{J+\delta(t)+[J-\delta(t)]\cos(k)\}\sigma_{x}-[J-\delta(t)]\sin(k)\sigma_{y}+\Delta(t)\sigma_{z}.
\label{eq:RMHamiltonian}
\end{equation}
Here $\sigma_{x},\sigma_{y}$ and $\sigma_{z}$ are three Pauli matrices.
$k\in[-\pi,\pi)$ is the quasimomentum along $x$-direction. The sign
of the parameter combination $J\delta_{0}\Delta_{0}$ controls phase
transitions in the system. The spectrum of $H(k,t)$ becomes gapless
when $J\delta_{0}\Delta_{0}=0$. When $2\pi t/T$ is viewed as a quasimomentum
along a synthetic dimensional, the Hamiltonian $H(k,t)$ is mapped
onto a two-dimensional lattice model~(also called the Harper-Hofstadter
model)~\cite{Harper1955,Hof1976}, describing the minimum realization
of a Chern insulator. At half-filling, this model is topologically nontrivial
with band Chern numbers $\pm1$. Thus the Hamiltonian $H(k,t)$ can
also be interpreted as a dynamical realization of the Harper-Hofstadter
model in a one-dimensional lattice, and the quantized Thouless pump
observed there is determined by the Chern number of the two-dimensional
parent model.

Experimentally, Thouless pump is observed by measuring the shift of
a wave packet center over an adiabatic cycle~\cite{Naka2016,Loh2016}.
To observe quantized pumping, the initial state is assumed to uniformly
fill the valence band, realizing a Wannier state in the lattice. However,
under more general experimental conditions, the initial state could
be out-of-equilibrium and coherently populates both the valence and
conduction bands. Then how will the interband coherence in the initial
state affect Thouless pump? In our recent studies, we found that for
nonequilibrium initial states with $k$-reflection-symmetric populations,
the Thouless pump is not quantized~\cite{Hailong2015,LongWen2015}.
This generalized Thouless pump contains two components, an integral
over the geometric Berry curvature weighted by initial band populations,
plus an accumulated dynamical effect due to interband coherence in
the initial state. The shift of wave packet center over an adiabatic
cycle is then given by~\cite{FOOTNOTE}
\begin{alignat}{1}
\Delta\langle x\rangle & =\Delta\langle x\rangle_{1}+\Delta\langle x\rangle_{2},\label{eq:Pump}\\
\Delta\langle x\rangle_{1} & =\frac{1}{2\pi}\sum_{n}\int_{-\pi}^{\pi}dk\int_{0}^{2\pi}d\beta B_{n,k}(\beta)\rho_{n,k}(0),\label{eq:Geometry}\\
\Delta\langle x\rangle_{2} & =-\frac{1}{\pi}\sum_{m,n,m\neq n}\int_{-\pi}^{\pi}dk{\rm Re}\left[C_{m,k}(0)C_{n,k}^{*}(0)W_{nm,k}(0)\right]\int_{0}^{1}ds\frac{\partial E_{n,k}(s)}{\partial k}.\label{eq:IBC}
\end{alignat}
where $m,n$ are energy band indices. $s\equiv t/T\in[0,1)$ is the
scaled time, $\beta=\beta(s)$ describes the adiabatic protocol (e.g.,
$\beta=2\pi s$ for a linear protocol). In Eq.~(\ref{eq:Geometry}),
$\rho_{n,k}(0)$ is the initial population of the $n$th energy band
at quasimomentum $k$. The Berry curvature $B_{n,k}(\beta)$ is defined
as
\begin{equation}
B_{n,k}(\beta)\equiv i\left[\left\langle \frac{\partial\psi_{n,k}(\beta)}{\partial\beta}\left|\frac{\partial\psi_{n,k}(\beta)}{\partial k}\right.\right\rangle -\left\langle \frac{\partial\psi_{n,k}(\beta)}{\partial k}\left|\frac{\partial\psi_{n,k}(\beta)}{\partial\beta}\right.\right\rangle \right],\label{eq:Curvature}
\end{equation}
where $|\psi_{n,k}(\beta)\rangle$ is an instantaneous eigenstate
of $H(k,\beta)$ with energy $E_{n,k}(\beta)$. In Eq.~(\ref{eq:IBC}),
$C_{n,k}(0)$ is the probability amplitude in Bloch basis $|\psi_{n,k}(0)\rangle$,
and thus the cross term $C_{m,k}(0)C_{n,k}^{*}(0)$ with $m\neq n$
captures the interband coherence in the initial state. The other function
$W_{nm,k}(0)$, also related to initial state coherence, is defined
as
\begin{equation}
W_{nm,k}(0)\equiv\left.\frac{\langle\psi_{n,k}(\beta)|\frac{\partial}{\partial\beta}|\psi_{m,k}(\beta)\rangle}
{i\left[E_{m,k}(\beta)-E_{n,k}(\beta)\right]}\frac{d\beta}{ds}\right|_{s=0}.
\label{eq:Wmnk}
\end{equation}
This quantity is sensitive to the switching-on speed $\left.\frac{d\beta}{ds}\right|_{s=0}$
of an adiabatic protocol. Moreover, it also has singular behaviors
around band touching points, and thus would be important around
topological phase transitions. Though the component $\Delta\langle x\rangle_{1}$
is already known in Thouless's original proposal, the contribution
of $\Delta\langle x\rangle_{2}$ to adiabatic transport is only discovered
recently in Ref.~\cite{Hailong2015}. This correction term is intriguing because
it is induced purely by interband coherence and
it is independent of the time duration $T$ of an adiabatic protocol.
Thus it is highly valuable to
experimentally observe this.  In the following sections, we will motivate such type of experiments
by studying how the coherence effects affect the pumping
with different kinds of specific initial states and adiabatic protocols.

\section{Adiabatic pumping of Gaussian states after quench\label{sec:Gaussian}}

\begin{figure}
\begin{center}

{%
  \includegraphics[scale=0.5]{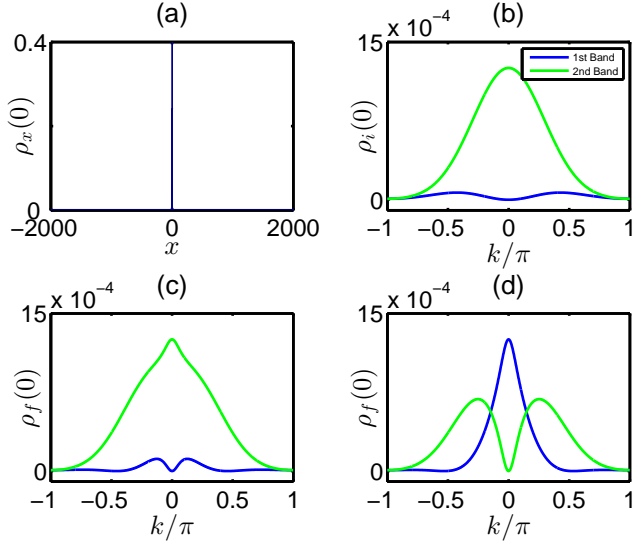}
}

\end{center}

\caption{(Color online)
(a) Profile of the Gaussian wave function in the position space and
(b) Probability distribution $\rho_{i}(0)$ on two bands along $k$ for the Gaussian state
before quenching at system parameters $\delta(0)=0.85$, $\Delta(0)=0$ and $J=0$. (c) Probability distribution $\rho_{f}(0)$ on two bands along $k$ of the Gaussian state after quenching the system parameter in the same topological phase. The final system parameter is $J_{f}=0.1$. (d) Probability distribution $\rho_{f}(0)$ of the Gaussian state after quenching the system parameter across phase transition to $J_{f}=-0.1$.\label{fig:gauinipop}}
\end{figure}

Quantum quench provides with us a general approach to prepare a state
with interband coherence. By starting with a state occupying only
a single band, and quenching one of the Hamiltonian's parameter to
another value, the state will coherently populate both the valence
and conduction bands of the post-quench Hamiltonian. The adiabatic
pumping of such a nonequilibrium initial state constitutes a generalized
Thouless pump, in which the geometric nature of adiabatic states
and accumulated non-adiabatic effects along a pumping path over a long time duration are
both important.

In this section, to study the effect of interband coherence in adiabatic
pumping, we propose to initialize the state of our system~(i.e., the time-dependent Rice-Mele
model discussed in the last section) with a Gaussian wave packet on the lattice.
In the lattice representation, the profile of the wave packet is given
by $Ae^{-(x-L/2)^{2}/d^{2}}$, where $A$ is the normalization constant.
In our simulation, we choose $L=4000$ and $d=20$. Such a parameter
choice is to ensure that the wave packet has a relatively broad distribution
in the momentum space, in order to capture interband coherence effects
efficiently. The other parameters for the pre-quench system
are chosen as $\delta(0)=0.85,\Delta(0)=0$ and $J=1$, closing
to the range of parameter values that can be realized experimentally~\cite{Naka2016}.
The probability distributions of
the Gaussian wave packet before the quench are shown in Figs.~\ref{fig:gauinipop}~(a)
and \ref{fig:gauinipop}~(b). We see that in momentum space, the Gaussian
wave packet mainly populates the conduction band.

To introduce interband coherence into the initial state, we quench
the system parameter $J$ suddenly from $J=1$ to different values
$J=J_{f}\in[-1,1]$. In the basis of the post-quench Hamiltonian,
the Gaussian wave packet occupies both the conduction and valence
bands at different values of $k$. In Figs.~\ref{fig:gauinipop}~(c)
and \ref{fig:gauinipop}~(d), we showed the probability distribution $\rho_{f}(0)$ of the
initial state in the basis of the post-quench Hamiltonian in two typical
cases. In Fig.~\ref{fig:gauinipop}~(c), we set $J_{f}=0.1$. Therefore the
post-quench and pre-quench Hamiltonians belong to the same topological
phase. In Fig.~\ref{fig:gauinipop}~(d), we choose $J_{f}=-0.1$, and thus the
post-quench Hamiltonian belongs to a different topological phase as
compared with the pre-quench Hamiltonian. Under such a quench, the
initial population of the two bands at $k=0$ get reversed. We give
an analytical explanation of this observation in Appendix.

\begin{figure}
\begin{center}

{%
  \includegraphics[scale=0.4]{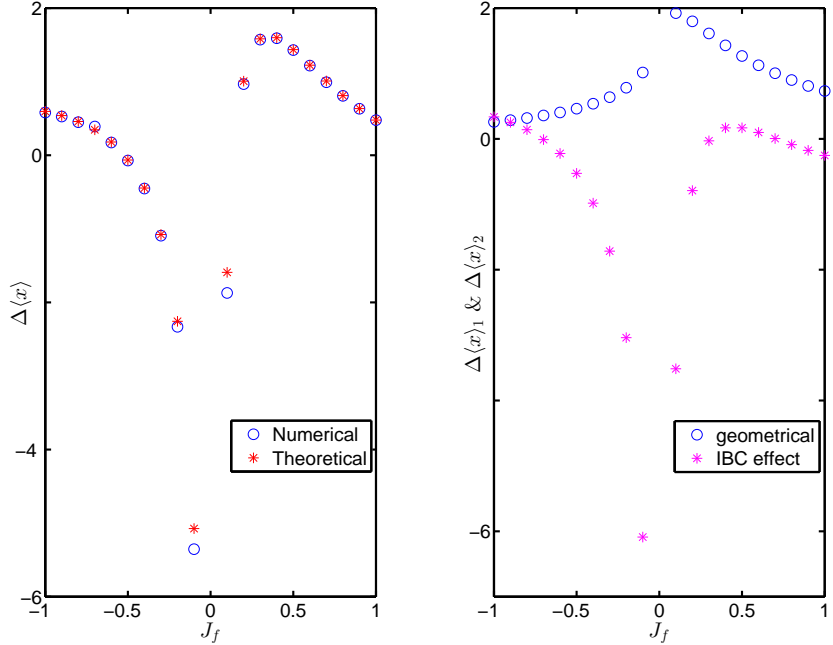}
}

\end{center}
\caption{(Color online) (a) Position expectation value after adiabatically
pumping the Gaussian state besides quenching the system parameter to
$J_{f}$. The numerical values~(blue circles) are in agreement
with theoretical values~(red stars) except at the small window
around the topological phase transition point, i.e., $J_{f}=0$.
The time duration of the linear pumping protocol is $T=1000$.
(b) The contribution of geometric and interband coherence~(IBC) parts to the change of
position expectation value, $\Delta\langle x\rangle_{1}$ and $\Delta\langle x\rangle_{2}$,
after quenching and adiabatically pumping the Gaussian state.
\label{fig:num vs theo}}
\end{figure}

Once an initial state with interband coherence is prepared, we can
pump it across the lattice under a given adiabatic protocol $\beta(s)$.
The adiabatic limit corresponds to taking the time duration $T\rightarrow\infty$.
In Fig.~\ref{fig:num vs theo}, we showed the shift of the center of post-quench
wave packets at different values of $J_{f}$ over an adiabatic cycle, under
a linear protocol $\beta(s)=2\pi s$. In Fig.~\ref{fig:num vs theo}~(a), The
numerical results~(in blue circles)~obtained from direct evolution
of Schrodinger equation is compared with the theoretical predictions~(in red stars)
of Eq.~(\ref{eq:Pump}). A good match is observed even if the post-quench
Hamiltonian has a very small gap~(i.e., around $J_{f}=0$). Moreover,
in Fig.~\ref{fig:num vs theo}~(b) we showed the contribution of weighted Berry
curvature $\Delta\langle x\rangle_{1}$ and interband coherence $\Delta\langle x\rangle_{2}$
to the shift of wave packet center $\Delta\langle x\rangle$ separately.
We observe that over the whole parameter domain $J_{f}$ of the post-quench
Hamiltonian, the pumping is dominated by the contribution of interband
coherence in $\Delta\langle x\rangle_{2}$. Especially, $\Delta\langle x\rangle_{2}$
is peaked at $J_{f}=0$, where the system undergoes a topological
phase transition. So we conclude that interband coherence in a nonequilibrium
initial state could not only play an important role in adiabatic transport,
but also be a powerful tool to detect topological phase transitions
dynamically. In the next section, we will further elaborate on these
points and highlight some other intriguing features of interband coherence
in adiabatic dynamics.

\section{Adiabatic pumping of Wannier states after quench\label{sec:Wannier}}

\begin{figure}
\begin{center}

{%
  \includegraphics[scale=0.5]{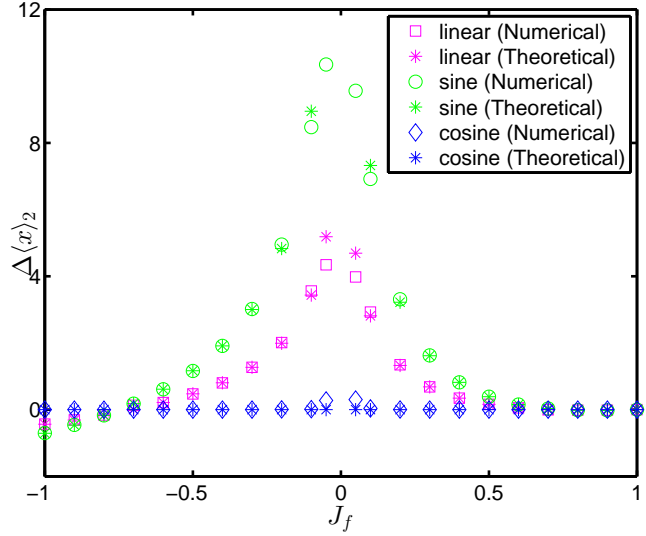}
}

\end{center}
\caption{(Color online) The contribution of interband coherence to the
shift of Wannier state under different pumping protocols.
Before the quench, the Wannier state was prepared at $J=1$,
occupying the lower band. The state was adiabatically pumped
after quenching the system parameter to $J_{f}$. Interband coherence
effect $\Delta\langle x\rangle_{2}$, from Eq.~(\ref{eq:IBC}), is shown for three different pumping protocols:
(i) $\beta_{L}=2\pi s$,
(ii) $\beta_{S}=2\pi\sin\left(\frac{\pi s}{2}\right)$, and
(iii) $\beta_{C}=2\pi[1-\cos(\pi s)]$, both numerically and theoretically
for $T=1000$. The $\beta_{S}$ pumping protocol has a larger contribution of interband
coherence effect compared to the other two protocols.
\label{fig:Difference-of-Interband}}
\end{figure}

In this section, we study adiabatic pumping of a Wannier state after
a quantum quench. Before the quench, the Wannier state is prepared
by superposing Bloch states uniformly across the valence band of the
pre-quench Hamiltonian at $J=1$. Such kind of states have been effectively prepared
in recent cold atom experiments~\cite{Naka2016,Loh2016}.
After the sudden quench, the Wannier state populates both the
conduction and valence bands of the post-quench
Hamiltonian at $J=J_{f}$ coherently. Starting with such an initial
state, we study how the contribution of interband coherence to
adiabatic pumping will be affected by selecting adiabatic protocols with
different switching-on speeds and time durations.

\begin{figure}
\begin{center}

\resizebox{0.6\textwidth}{!}{%
  \includegraphics[scale=0.5]{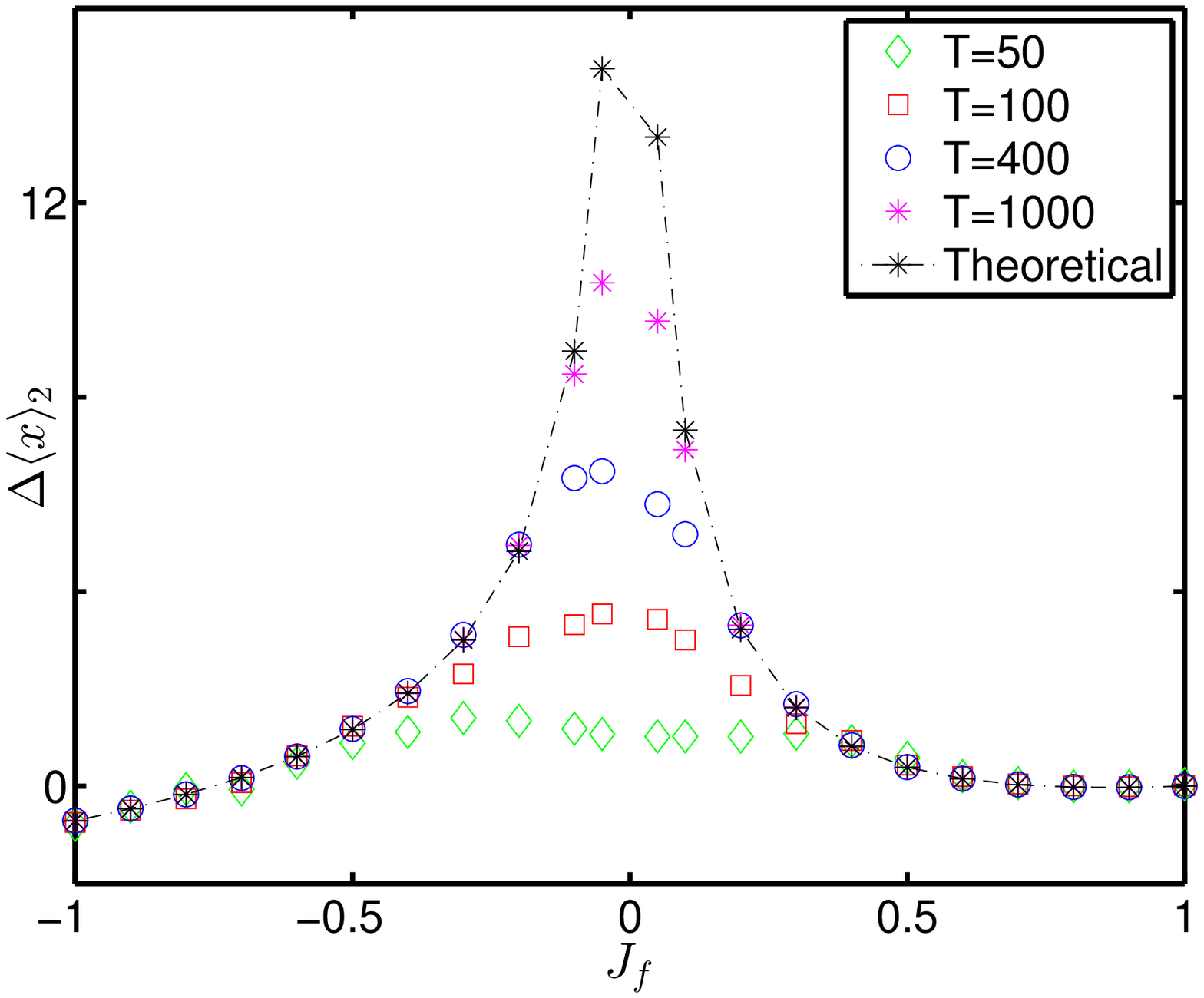}
}

\end{center}

\caption{(color online) Interband coherence effect for $\beta_{S}=2\pi\sin\left(\frac{\pi s}{2}\right)$ pumping protocol with different time durations $T$. Before the quench, the system parameter $J=1$. Interband coherence effects for pumping durations $T=50$ (diamond), $T=100$ (square), $T=400$ (circle), and $T=1000$ (star) are similar,
except at the window around phase transition point.}
\label{fig:IBCWanprotocols}
\end{figure}

First, we explore the switching-on behavior of the adiabatic driving
field by considering three different protocols:
(i) $\beta_{L}=2\pi s$, with a switching-on speed $2\pi$,
(ii) $\beta_{S}=2\pi\sin\left(\frac{\pi s}{2}\right)$,
with a switching-on speed $\pi^{2}$, and
(iii) $\beta_{C}=2\pi[1-\cos(\pi s)]$,
with a switching-on speed $0$. Based on our theoretical prediction
in Eq.~(\ref{eq:IBC}), the protocol $\beta_{S}$ has a larger switching-on
speed, and thus introducing larger interband coherence effects to
adiabatic pumping than the other two protocols. Note also that the
protocol $\beta_{C}$ has a vanishing switching-on speed,  thus
suppressing the contribution of interband coherence in adiabatic pumping.
In Fig.~\ref{fig:Difference-of-Interband}, we compare our theoretical predictions with direct
numerical simulations, and obtain nice agreement between theory and numerics. In particular, in experiments,
the difference in pumping among the three different adiabatic protocols can be regarded as a clear signature of
the above-mentioned coherence-induced correction.
Similar to the pumping of a Gaussian state after quench, we also observed
a peak in the interband coherence induced wave packet shift $\Delta\langle x\rangle_{2}$
around the topological phase transition point $J_f=0$. Such a universal
behavior beyond specific initial state preparations again highlights
the possibility of using interband coherence as a tool to detect topological
phase transitions. The sensitivity of interband coherence effects
to pumping protocols also has potential quantum control applications in adiabatic processes.

Since the contribution of interband coherence to adiabatic pumping
originates from the accumulation of dynamical effects over a long
time duration, it is important to know whether and how will it be
affected if the time duration of the adiabatic process is changed.  Our theory indicates that it will be independent of $T$, but that
is expected to be true only if the actual time duration of the pumping protocol is sufficiently long.
Let us focus on the adiabatic protocol $\beta_{S}=2\pi\sin\left(\frac{\pi s}{2}\right)$,
and pump the initial state over adiabatic cycles with different
time durations $T$. The results are shown in Fig.~\ref{fig:IBCWanprotocols}. We
observe that the numerical simulations fit our theory quite well even
for relatively small values of $T$, provided that the post-quench Hamiltonian
is away from the gapless point $J_{f}=0$. Close to $J_{f}=0$,
the gap of the post-quench Hamiltonian becomes small, and strong non-adiabatic
effects appear in the dynamics. It is then anticipated that our theoretical
predictions will deviate from numerical results, since the former
is derived under first order adiabatic approximations. However, for
not-too-short pumping durations, the peak around the topological phase
transition point is still well-captured by the contribution from interband
coherence.   Thus we conclude that for nonequilibrium initial states,
the interband coherence predicted by our theory
can be observed even when the adiabatic protocol is not very slow.  This observation is really good news for possible experiments.

\section{Summary and outlook\label{sec:Summary}}

In this work, we performed detailed theoretical and numerical analysis on Thouless
pump with nonequilibrium initial states, based on cold-atom experiments that are already available.
It was shown that initial
states with interband coherence can be prepared from Gaussian or Wannier
states through quantum quenches. The change of position expectation
value using such initial states~(which can be measured in available experiments)
during an adiabatic cycle is composed
of two parts with different physical origins. Though both components
converge to finite values in the adiabatic limit, one of them comes from
geometric Berry curvature of the energy band, while the other
arises from the accumulation of interband coherence effects over a pumping cycle.
The contribution of interband coherence to Thouless pump is sensitive to the switching-on
behavior of an adiabatic protocol and the size of band gap, thus having
potential applications in designing adiabatic control strategies and
detecting topological phase transitions. It should be promising and
feasible to verify our predictions using  existing experimental
setups for Thouless pump~\cite{Naka2016,Loh2016}. Such experiments will be useful towards
extensions of Thouless pump and a better understanding of quantum adiabatic transport.

\begin{acknowledgments}
Naresh would like to thank Prof.~Dario Poletti for helpful discussions. This work is supported by NRF Grant No. NRF-NRFI2017-04 (WBS No. R-144-000-378-q281) and by Singapore Ministry of Education Academic Research Fund Tier I~(WBS No. R-144-000-353-112).  
\end{acknowledgments}


\appendix
\vspace{1cm}

\section{Population reversal at $k=0$}

In this appendix, we show that the population of Gaussian state at quasimomentum $k=0$ is reversed if the pre-quench
and post-quench Hamiltonians belong to different topological phases.

In momentum space, the Hamiltonian of Rice-Mele model is given by Eq.~(\ref{eq:RMHamiltonian}),

\begin{eqnarray*}
h(k,t)=\{J+\delta(t)+[J-\delta(t)]\cos(k)\}\sigma_{x}-[J-\delta(t)]\sin(k)\sigma_{y}+\Delta(t)\sigma_{z}.
\end{eqnarray*}

The instantaneous energy spectrum $E(k,t) $is then given by:

\begin{eqnarray}
E(k,t)  & = & \sqrt{\Delta^{2}(t)+\{J+\delta(t)+[J-\delta(t)]\cos(k)\}^2+[J-\delta(t)]^{2}\sin^2(k)}
\nonumber \\
 & = &  \pm\lambda
\end{eqnarray}

\begin{figure}
\begin{center}
  \includegraphics[scale=0.5]{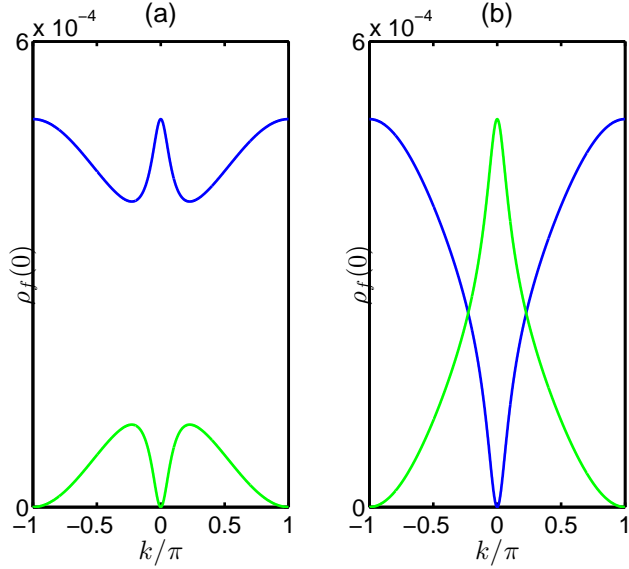}
\end{center}
\caption{(Color online)  The initial system parameter
$J=1$ and $t=0$ and topological
phase transition occurs at the point $J=0$. (a) $\rho_{f}(0)$
on two bands along $k$, after quenching the system to the same phase. The final system parameter $J_{f}=0.1$. The probability distribution of the first band is exclusively populated at the value
$k=0$ as is the case before the quench. (b) $\rho_{f}(0)$
on two bands along $k$, after quenching the system to a different phase. The final system parameter $J_{f}=-0.1$. Note
that the probability distribution of the second band is now exclusively
populated at the value $k=0$ which is in contrast
to the case before the quench.\label{fig:Initial-state.}}
\end{figure}

And the eigenvectors are,

\begin{eqnarray}
|\psi_{1}\rangle & =\frac{1}{\sqrt{\alpha^{2}+\omega^{2}+(\lambda+\beta)^{2}}} & \left(\begin{array}{c}
\alpha-i\omega\\
-\lambda-\beta
\end{array}\right)\
\label{eq:psi1}
\end{eqnarray}

\begin{eqnarray}
|\psi_{2}\rangle & =\frac{1}{\sqrt{\alpha^{2}+\omega^{2}+(\lambda-\beta)^{2}}} & \left(\begin{array}{c}
\alpha-i\omega\\
\lambda-\beta
\end{array}\right)
\label{eq:psi2}
\end{eqnarray}

Here, $\alpha = J+\delta(t)+[J-\delta(t)]\cos(k)$, $\beta=\Delta(t)$, and $\omega=-[J-\delta(t)]\sin(k)$.

Let us now prepare our initial state $|\psi\rangle$ by populating
only the first band in $k$-space at a particular value of $J=J_{i}$. Then we quench the state by changing the parameter $J$ of the Hamiltonian to $J=J_{f}$.

The wave function in the pre-quench eigenbasis, is given
as,
\begin{eqnarray}
|\psi\rangle & = & |\psi_{1}\rangle_{i}\label{eq:bqstate}
\end{eqnarray}

And in the post-quench eigenbasis it is given as,

\begin{eqnarray}
|\psi\rangle & = & \eta_{1}|\psi_{1}\rangle_{f}+\eta_{2}|\psi_{2}\rangle_{f}
\end{eqnarray}

at each $k$. The two coefficients $\eta_1$ and $\eta_2$ are given by

\begin{eqnarray}
\eta_{1} & = & _{f}\langle\psi_{1}|\psi_{1}\rangle_{i}\\
\eta_{2} & = & _{f}\langle\psi_{2}|\psi_{1}\rangle_{i}.
\end{eqnarray}

Note that $\alpha$, $\omega$ and $\lambda$ are all depend on the system parameter $J$. Using Eqs.~(\ref{eq:psi1}) and (\ref{eq:psi2}), we obtain

\begin{eqnarray}
\eta_{1} & = & \frac{(\alpha_{i}-i\omega_{i})(\alpha_{f}+i\omega_{f})+(\lambda_{i}+\beta)(\lambda_{f}+\beta)}{\sqrt{\alpha_{i}^{2}+\omega_{i}^{2}+(\lambda_{i}+\beta)^{2}}\sqrt{\alpha_{f}^{2}+\omega_{f}^{2}+(\lambda_{f}+\beta)^{2}}}\\
\eta_{2} & = & \frac{(\alpha_{i}-i\omega_{i})(\alpha_{f}+i\omega_{f})-(\lambda_{i}+\beta)(\lambda_{f}-\beta)}{\sqrt{\alpha_{i}^{2}+\omega_{i}^{2}+(\lambda_{i}+\beta)^{2}}\sqrt{\alpha_{f}^{2}+\omega_{f}^{2}+(\lambda_{f}-\beta)^{2}}}
\label{eq:psi-final}
\end{eqnarray}

For simplification, let us look at the particular value of $k=0$ at $t=0$, for different values of $J$.
So, we have $\delta = \delta_{0}$, $\Delta = 0$, $\alpha=2J, \beta=0, \omega=0, \lambda=|\alpha|= 2|J|$.
Therefore, when the pre-quench and post-quench systems belong to the same topological phase, we have $J_{i}J_{f}>0$. It implies that

\begin{eqnarray}
\eta_{1}|_{k=0} & = & \frac{J_{i}J_{f}+J_{i}J_{f}}{\sqrt{J_{i}^{2}+J_{i}^{2}}\sqrt{J_{f}^{2}+J_{f}^{2}}}=1.
\end{eqnarray}

So the population of the lower band is preserved at $k=0$, as shown in Figure~\ref{fig:Initial-state.} (a).
But if we quench our Hamiltonian to the other topological phase, we will have $J_{i}J_{f}<0$. Then Eq.~(\ref{eq:psi-final}) implies that

\begin{eqnarray}
\eta_{1}|_{k=0} & = & \frac{J_{i}J_{f}-J_{i}J_{f}}{\sqrt{J_{i}^{2}+J_{i}^{2}}\sqrt{J_{f}^{2}+J_{f}^{2}}}=0;
\end{eqnarray}

Therefore the population at $k=0$ in the lower band is fully transferred to the higher band at the same $k$ after the quench. The result is shown in Fig.~\ref{fig:Initial-state.} (b).

So we conclude that the population at $k=0$ is reversed
if the pre-quench and post-quench Hamiltonians belong to
different topological phases.

\end{document}